\documentclass[11pt]{article}
\usepackage{graphicx}

\begin{document}

\hoffset=-20mm
\voffset=-8pt

\vspace*{26mm}
\begin{figure*}[t]		
\begin{center}
\includegraphics[width=17.5cm, angle=0]{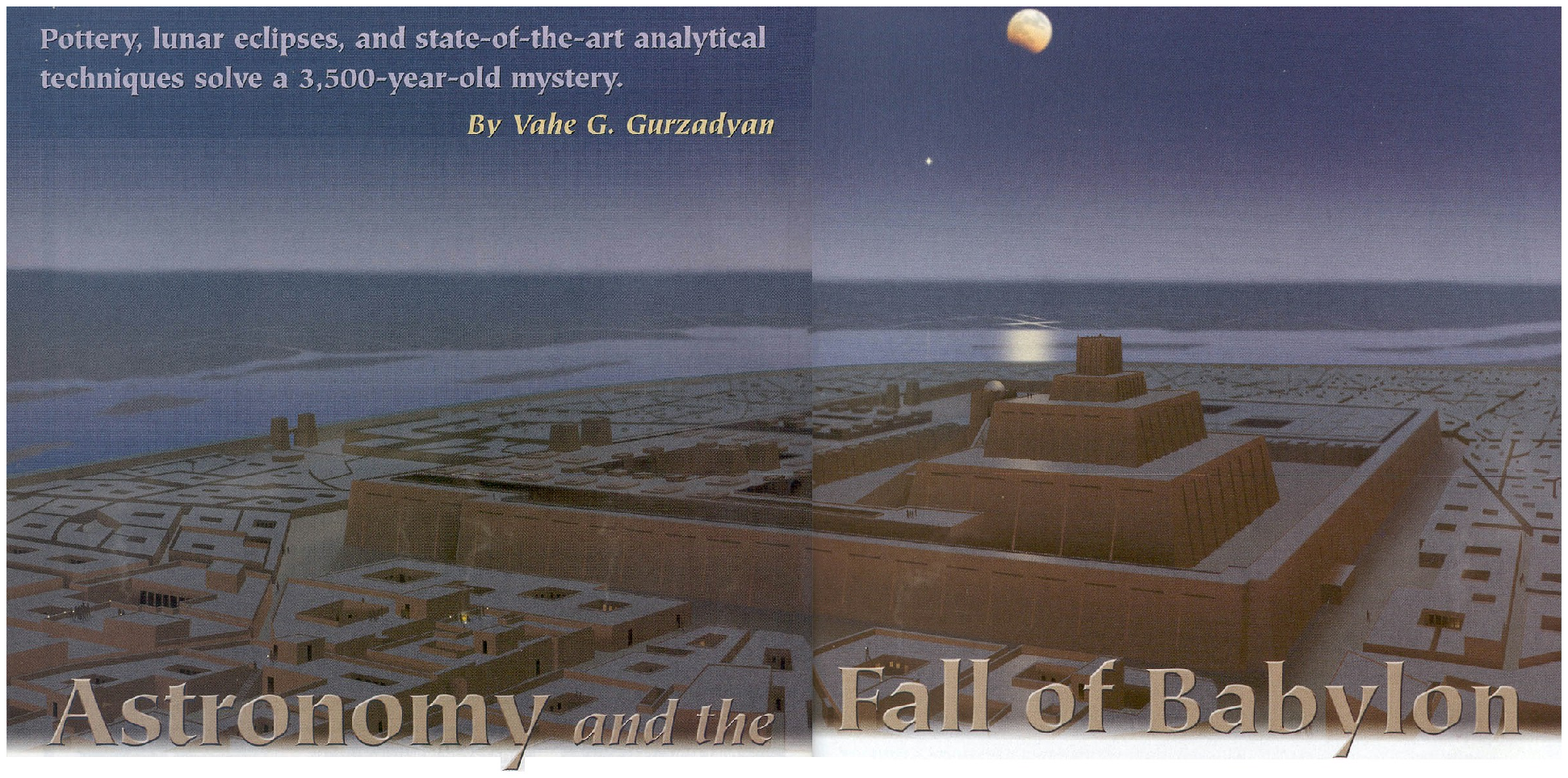}
\end{center}
\end{figure*}

{\Large\bf
\noindent
Astronomy and the Fall of Babylon\\[3mm]
}\\[6mm]
{\large V.G.Gurzadyan}\\[6mm]
{Yerevan  Physics Institute, Armenia and \\
ICRA,  University  of  Rome ``La Sapienza", Italy}\\[5mm]

\noindent(Published in {\it Sky \& Telescope},  v.100, No.1 (July), pp. 40-45, 2000; 
design credit:{\it Sky \& Telescope}.
)	 

\vspace{0.1in}

\section{Introduction}

It's not often when sophisticated techniques developed for astronomy can 
answer an earthly mystery that has persisted for thousands of years. Yet 
there is a direct link joining the Cosmic Background Explorer (COBE) 
spacecraft and lunar laser-ranging to the precise dating of a celebrated 
historical event - the fall of Babylon to the Hittites in the second 
millennium B.C. 

One of the most famous cities in the ancient world, Babylon was 
strategically located on the Euphrates River. There it wielded political 
power and controlled trade throughout a large region of Mesopotamia 
(modern-day Iraq). Yet we remember it today as a fount for our 
scientific 
heritage. Babylonian astronomy is directly echoed in the Almagest of 
Claudius Ptolemy (about A.D. 140), which epitomized this science until 
the time of Copernicus 14 centuries later. Even nowadays our culture is 
bound to such inventions as the sexagesimal system and the zodiac. 

My involvement with Babylonian astronomy started in 1995, when I met 
Hermann Gasche, a leading European archaeologist, coordinator of 
excavations in various areas of Mesopotamia, and author and editor of 
many books on the archaeology of the region. Our association seemed 
preordained. 
Several years earlier, while visiting Armenia, Gasche became aware of 
unpublished works prepared in the 1920s and '30s by the archaeologist 
Ashkharbek Kalantar, my grandfather, and he arranged for their 
publication. 

\begin{figure}[t]		
\begin{center}
\includegraphics[width=9.88cm]{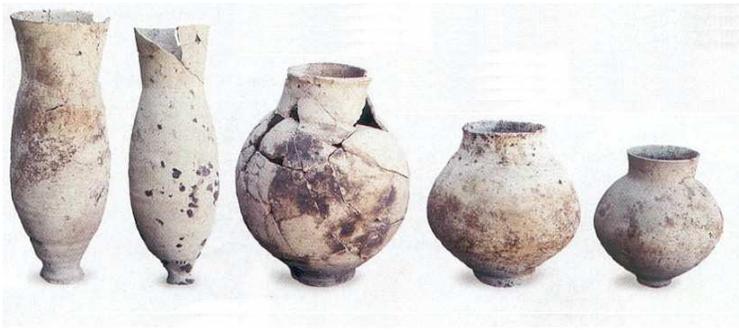}
\end{center}
\caption{
One key for establishing dates in the ancient world is the changing 
shapes 
of pottery. These  vessels are typical for the years just before 
Babylon's 
fall to the Hittites. The tall ones are probably mugs for beer, a 
favorite 
Babylonian beverage.}
\end{figure}

As we sat in his Paris studio full of pictures of pottery, Gasche 
excitedly 
told me about a problem he had been working on with two American 
researchers, archaeologist James Armstrong and Assyriologist Steven 
Cole. The vessel shapes he had studied at various archaeological sites 
couldn't 
be accommodated within the time frame of the so-called Middle Chronology 
(see box) of Mesopotamia, which was the most commonly accepted scheme 
but 
also the most criticized. (Archaeologists frequently use pottery 
evidence 
to help date ancient cultures.) To reconcile what Gasche and his 
colleagues 
had found required shortening the Middle Chronology by about a century. 

Part of the reason behind Gasche's excitement was a group of tablets 
recently unearthed at a site in a Baghdad suburb called Tell Muhammad. 
Among these tablets were two bearing references to a lunar eclipse that 
occurred 38 years after Babylon was resettled. He wanted to see whether 
I could use that eclipse and other astronomical sources to establish an 
absolute chronology for around the time Babylon fell and perhaps 
simultaneously resolve the pottery enigma. Such a chronology can only be 
established through astronomical records like those on the eclipse 
tablet. 
Indeed, the standard Near Eastern chronology from about 1400 B.C. to 700 
B.C. - based on the so-called Assyrian Kinglist - is anchored by records 
of the solar eclipse of June 15, 763, B.C. 

My first step was to carefully examine the existing literature and the 
methodology of ancient Babylonian astronomers and mathematicians. It 
soon 
became evident that chronologies for eras earlier than 1400 B.C. - based 
on 
the so-called Venus Tablet from the reign of King Ammisaduqa of Babylon 
- were flawed. The reason is trivial: although the tablet contains 
sightings 
of the planet that could conceivably be dated, its 20 identified 
fragments 
are actually corrupted copies made about a thousand years after the 
events took place. 

The Venus Tablet is one (Number 63) of a series of some 70 tablets 
collectively known as Enuma Anu Enlil. Two others (Numbers 20 and 21) 
record a pair of lunar eclipses connected with the Third Dynasty of Ur, 
which dominated Babylonia around 4,000 years ago. 

\section{Unscrambling the Evidence}
 
I started with a reexamination of the Venus Tablet. This remarkable 
text, 
identified in the 1910s by the Jesuit Franz Kugler, contains data on 
first 
and last visibilities of Venus during a 21-year period that is believed 
to involve the reign of King Ammisaduqa, who governed a little over a 
century after the famous Babylonian king Hammurabi. 

\begin{figure}[t]		
\begin{center}
\includegraphics[height=10cm]{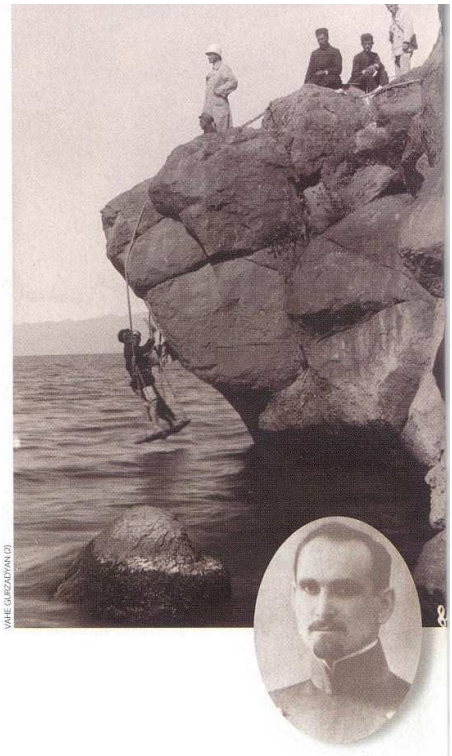}
\end{center}
\caption{
Copying cuneiform texts isn't always easy! In 1926 Ashkharbek Kalantar, 
the 
author's grandfather (in helmet), supervised the work involving a 
Urartian 
rock inscription at lake Sevan, the largest in Armenia. Today the water 
level is 20 meters lower. 
}
\end{figure}

\begin{figure}[t]		
\begin{center}
\includegraphics[width=7.06cm]{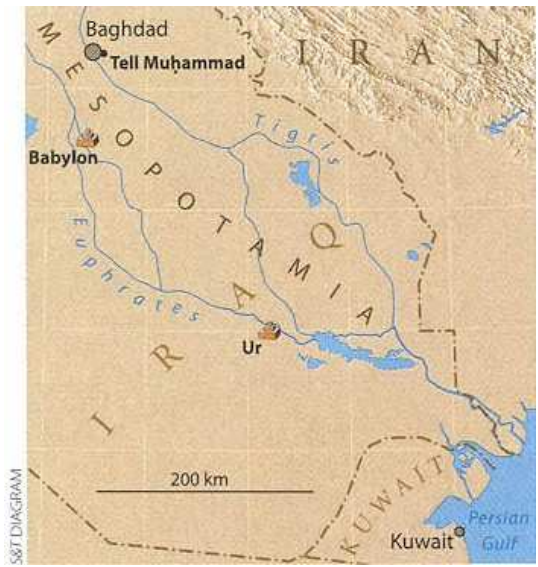}
\end{center}
\caption{
Ancient sites mentioned in the text are included on this map, which also 
shows modern Baghdad. }
\end{figure}

Following an orbit inside Earth's, Venus becomes temporarily invisible 
when it aligns itself either between the Earth and Sun (inferior conjunction) 
or beyond the Sun (superior conjunction). These passages mark when Venus 
switches from being a Morning Star to an Evening Star or vice versa. 
Very important to the Babylonian astronomers, as documented in the Venus 
Tablet, 
is that these morning/evening visibility cycles repeat almost precisely 
in the sky every 8 years. (Five of Venus's conjunction-to-conjunction 
periods of 584 days equal 8 years within a couple of days.) 

Almost from the time of its discovery it became evident that the Venus 
Tablet contains errors made by the original scribe as well as later 
copyists. Also, parts of it are obscured and unreadable. So it's not 
surprising that scholars have disagreed on its usefulness for dating 
events 
in the ancient Babylonian world. Erica Reiner and David Pingree, for 
example, claimed it was impossible to extract reliable chronological 
information from the tablet. Nevertheless, researchers have continued to 
rely on the Venus Tablet to generate chronologies for the centuries 
preceding the fall of Babylon. 

In modern jargon, we can say the tablet has "noisy" data, and that's 
where 
my experience with COBE proved useful. In 1996 I worked with COBE team 
member Sergio Torres on ways to pluck real signals from COBE's complete 
data set. This was no easy task because of pervasive contamination by 
noise 
from the cosmos, the Earth's atmosphere, and the detectors themselves. 
We 
were analyzing the distortion of "hot" and "cold" spots in sky maps of 
the 
cosmic microwave background, and we finally detected a tiny signal that 
matched a theoretical prediction to suggest the universe might have 
negative curvature and expand forever. (We published the result in 
Astronomy and Astrophysics, Vol. 321, page 19, 1997.) 
  
When I applied the same kind of sifting to the Venus Tablet, I came up 
empty-handed. The Monte Carlo technique of random sampling, as well as 
other statistical schemes, revealed noise but no significant signal 
except for that of the 8-year Venus cycle. 

Another lesson from COBE is that the nature of the noise itself can be 
studied - in other words, some corruptions might be identified because 
they have a systematic character. The opportunity to apply such modern 
analytical techniques to archaeology is new and might lead to important 
insights about ancient texts. 

\begin{figure}[t]		
\begin{center}
\includegraphics[width=4.5cm]{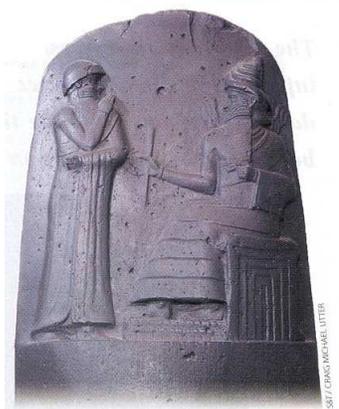}
\end{center}
\caption{
Hammurabi is the most recognizable name pertaining to ancient Babylon. 
The 
bas-relief at the top of this stele depicts the king (left) receiving 
his 
famous code of laws from the Sun-god Shamash. Hammurabi ruled in the 
17th 
century B.C. according to the new chronology described in this article. 
This stele, a copy of one in the British Museum, is at Harvard 
University's 
Semitic Museum.}
\end{figure}

\begin{figure}[t]		
\begin{center}
\includegraphics[width=8cm]{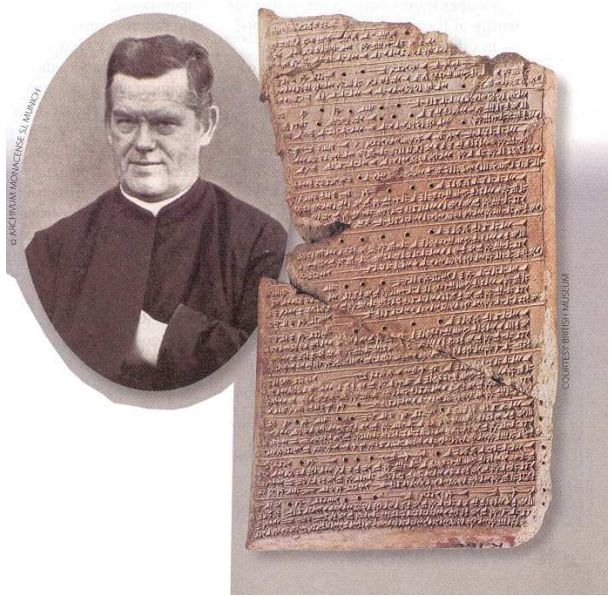}
\end{center}
\caption{
This tablet with its cuneiform text is believed to record 16th-century 
B.C. 
visibilities of the planet Venus and was long regarded as providing a 
key 
to dating events in ancient Babylon. However, the author has shown that 
its 
data are corrupted and unusable. 
The so-called Venus Tablet, which has confounded many Babylonian 
scholars, 
was first recognized as containing chronological information by the 
Jesuit 
priest Franz Xaver Kugler in the 1910s. Credit: Archivum Monacense SJ, Munich. 
}
\end{figure}

\begin{figure}[t]		
\begin{center}
\includegraphics[width=16cm]{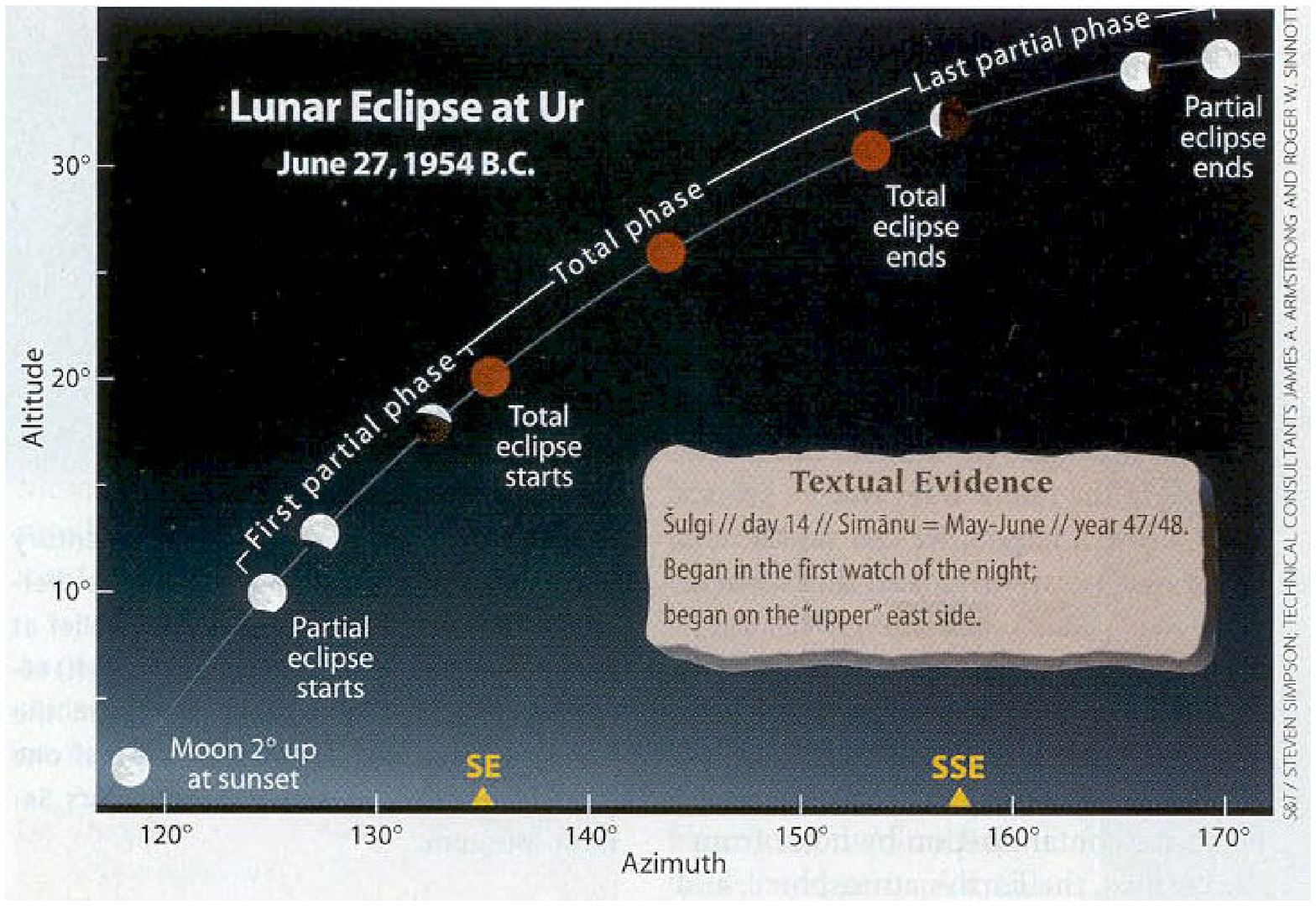}
\end{center}
\caption{
The principal phases of the June 27, 1954, B.C. total lunar eclipse are 
depicted here together with their Universal Times. }
\end{figure}

Then I turned to dating the two lunar eclipses recorded during the Third 
Dynasty of Ur, which occurred more than 400 years before Babylon's fall. 
The problem was to accurately match up modern predictions of what should 
have been seen with what was actually observed. At first glance this 
seems to be a simple task. Indeed, there is plenty of software for home 
computers that can do the calculations, and there are lists that tell where 
eclipses occurred thousands of years ago and give such circumstances as their 
beginning and ending times to an accuracy of minutes. 

However, before utilizing such resources one has to clearly understand 
the 
approximations inherent in both the input data and in the algorithms for 
the calculation. A program can work well over short time spans but not 
over long ones, because planetary motions are essentially nonlinear and 
initial 
errors propagate exponentially. Therefore, for an orbit to be precisely 
defined over a given span of time, the input data have to be 
sufficiently 
accurate and all planetary perturbations have to be properly taken into 
account. The perturbations, particularly, are responsible for many 
unforeseen effects, including chaos and other unpredictables. 

Archaeologists are basically interested in two consequences of such 
calculations. One is the determination of the moment the Moon arrives at 
a point in its orbit opposite the Sun, when it can pass through the 
Earth's 
shadow and be eclipsed. The second is to find the local time an 
observer's 
clock would show for the event. The physical interrelated effects that 
contribute to uncertainty include tidally and nontidally induced 
variations 
in the Earth's rotation rate and the gradually increasing distance of 
the 
Moon from Earth. With events that happened 4,000 years ago, like the Ur 
eclipses, these effects can result in prediction errors of up to two 
hours, not merely minutes, despite using the most accurate data 
available.\footnote{My work owes a debt to visionaries at the dawn of space flight. 
Historically, lunar occultations of stars provided information about the 
Moon's position. But a dramatic improvement in the accuracy of the data 
occurred after Apollo astronauts placed laser reflectors on the Moon. 
Kenneth Nordtvedt, the initiator of the project, told me how he once 
deliberately took the same plane as physicist Robert Dicke, then 
chairman 
of NASA's Physical Science Committee, so he could convince him to 
include 
the laser-ranging reflectors on forthcoming Apollo missions. As a 
result, 
we now have much better knowledge of the parameters that define the 
Moon's 
orbit, its principal perturbations by the Sun, Venus, and Jupiter, and 
other factors.} 

My search was constrained by the textual evidence in Tablets 20 and 21 
of 
the {\it Enuma Anu Enlil}. The two eclipses described there had been linked by 
ancient astrologers with decisive events in Babylonian history: the 
death 
of the greatest of the kings of Ur, Sulgi, and the destruction of Ur at 
the 
hands of the Elamites. The ancient scribes preserved important 
descriptive 
information about these eclipses that is crucial for dating them 
correctly: 
the time of day that each began and where the Moon was in the sky. We 
also 
know that the eclipses had to occur 41 to 44 years apart because we know 
the {\it relative} dates of the two historical events to which they were 
connected. 

My colleagues and I agreed that I should attempt to identify the two 
eclipses by scanning a 300-year interval centered on date of the fall of 
Ur (2005 or 2004 B.C., according to the Middle Chronology). As it turned 
out, 
across these three centuries only the eclipses of June 27, 1954, B.C. 
and 
March 16, 1912, B.C. fit the ancient descriptions of the eclipses within 
possible inaccuracies of interpretation. 
  
\begin{figure}[t]		
\begin{center}
\includegraphics[width=4cm]{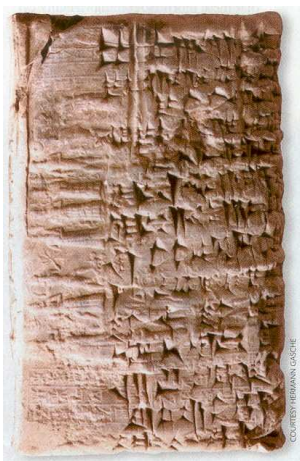}
\end{center}
\caption{
At the left of this Babylonian sales contract are sealings that 
serve as signatures of witnesses. Seals found at the ancient 
Syrian city of Terqa can be made self-consistent in time only
through the chronology described here.
}
\end{figure}

\section{Assembling the Case}

Now we could fix a date for the fall of Babylon. We had the date of the 
fall of Ur, and the ancient records were detailed enough for us to 
establish, within only a few years, the length of time that separated 
Ur's 
collapse from events that took place in Babylon several centuries later. 
Next, the Venus Tablet allowed us to link the reign of Ammisaduqa and 
the 
8-year cycle of Venus. Only simple arithmetic was needed to identify the 
possible Venus-derived dates for the fall of Babylon. When those were 
compared with the information from the lunar eclipses, the date of 
Babylon's fall could be fixed as 1499 B.C. This date, some 96 years more 
recent than the Middle Chronology date of 1595 B.C., fits well with the 
pottery evidence that led Gasche to ask me to look at the ancient 
astronomical records in the first place. 

Having established this date, I could now look at the information about 
the 
eclipse as recorded on the tablets from Tell Muhammad and determine when 
Babylon was resettled. However, in contrast with the Enuma Anu Enlil 
record, I was not dealing with eclipse descriptions. The Tell Muhammad 
tablets simply mention the eclipse in a so-called year-name that can be 
translated as "The year that the Moon was eclipsed." The two tablets 
also 
bear a second date formula as well: "Year 38 that Babylon was 
resettled." 

\begin{figure}[t]		
\begin{center}
\includegraphics[width=16cm]{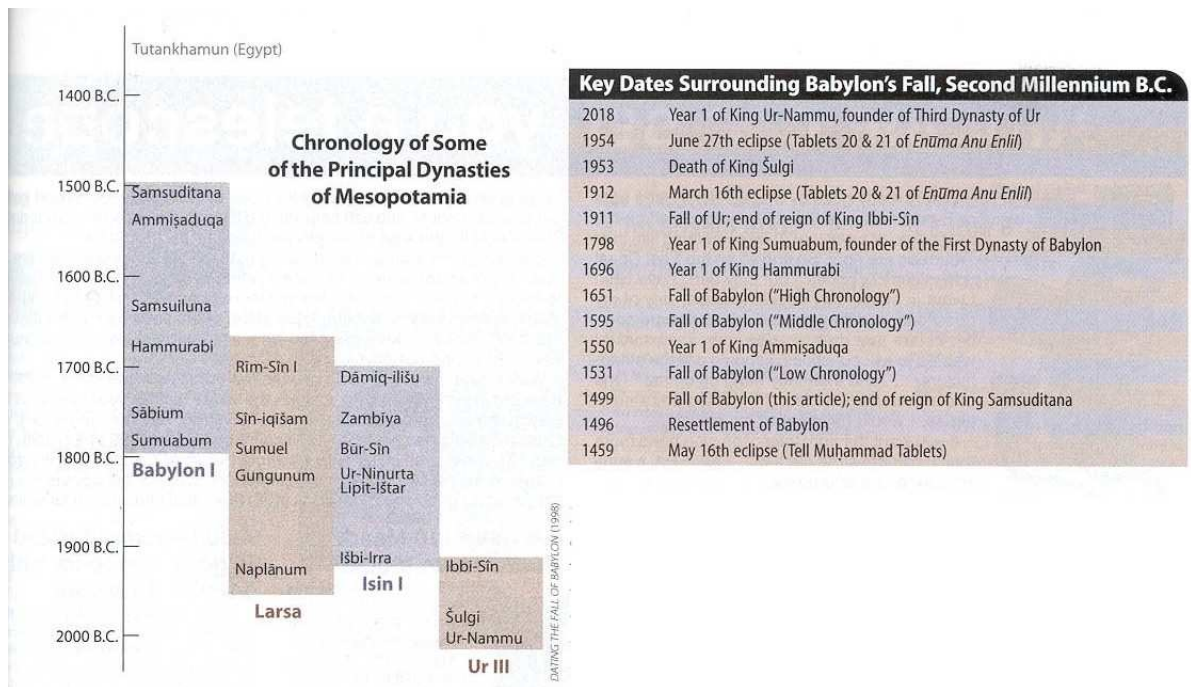}
\end{center}
\caption{
Chronology of Some of the Principal Dynasties of Mesopotamia
}
\end{figure}

\begin{figure}[t]		
\begin{center}
\includegraphics[width=4cm]{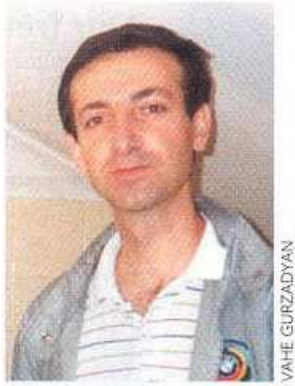}
\end{center}
\caption{
 Vahe G. Gurzadyan, author of articles and books on stellar dynamics and 
mathematical methods in cosmology, works at the Yerevan Physics 
Institute 
and Garni Space Astronomy Institute in Armenia as well as at the 
International Center of Relativistic Astrophysics, University of Rome 
"La 
Sapienza," Italy. The technical details of the study reported here are in 
the book {\it Dating 
the Fall of Babylon (Mesopotamian History and Environment)}, Series II, 
Memoirs IV, University of Ghent and the Oriental Institute of the 
University of Chicago, 1998.
}
\end{figure}

 It is unusual for texts to be dated according to two different systems. 
What is probably going on with these and other Tell Muhammad tablets is 
that an old, local year-name dating system was being gradually replaced 
by 
a new, Babylon-oriented one, and during the changeover the scribes used 
both kinds of dates. In any case, this redundancy allows us to date the 
resettlement of Babylon after its fall to the Hittites. Based on the 
available evidence, the Tell Muhammad eclipse most likely occurred on 
May 
16, 1459, B.C., so Babylon was resettled in 1496 B.C., only three years 
after its collapse.\footnote{In Babylon, each year was named for 
something noteworthy that occurred, 
such as a military victory or building project. However, this name was 
applied to the year following the memorable event. Thus, the so-called 
name-year, "the year of the eclipse," referring to the May 16, 1459, 
B.C. eclipse, is 1458 B.C.}

\section{Epilogue}
 
Our absolute chronology follows work by generations of scholars. This 
quest 
has been important because any change in the Babylonian chronology 
affects 
dates of events in other ancient kingdoms of the Near East. Elamite 
dynasties, the Old Hittite Kingdom, the Levant in the Middle Bronze Age, 
and the Second Intermediate Period in Egypt must now be fitted into this 
new scheme. 

Subsequent to our study, my colleagues and I learned of independent 
investigations that strongly support our new chronology. In particular, 
Guido Gualandi, who has studied the seals from the ancient city of Terqa 
in 
eastern Syria, reports that only by using a "low" chronology like ours 
can 
he make sense of the similarities among seals of the city's different 
kings. These similarities are simply too great to be easily explained if 
the rulers involved are separated from one another by spans of time as 
long 
as those necessitated by the traditional Middle or High Chronologies. 
Furthermore, studies of records in Egyptian papyruses from around 1800 
and 
1500 B.C. by Rolf Krauss also support our results rather than the Middle 
Chronology, which Krauss says nobody ever really believed! This 
prominent 
scholar confessed to me that he had been almost certain that the 
absolute 
chronology of the Near East would not be determined during his lifetime. 

Best of all, the morning after we had established the date of the fall 
of 
Babylon, I visited the British Museum with my daughter, Diana. The 
Babylonian kings there were smiling at us!

\section{Babylonian Chronologies} 

\begin{figure*}[t]		
\begin{center}
\includegraphics[width=4.23cm]{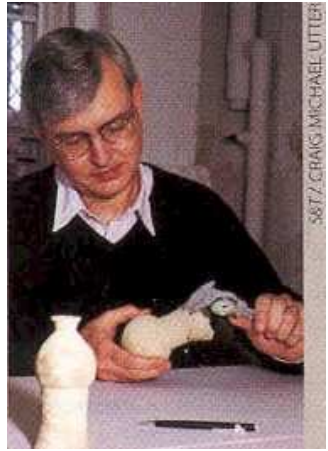}
\end{center}
\caption{
James Armstrong
}
\end{figure*}

Comments by James A. Armstrong 
\vspace{0.2in}  

Using ancient records, historians can reconstruct the Babylonian 
chronology 
very precisely back to 500+ years before Babylon's defeat at the hands 
of 
the Hittite king Mursili I. However, providing absolute dates over this 
half-millennium span has proved to be very difficult, particularly 
because 
it is separated from the earliest reliable Mesopotamian dates, which 
cluster around 1400 B.C., by an intervening period of unknown length. 

Since simply "counting back" from 1400 B.C. is not possible, researchers 
have turned to the Venus Tablet for help (see text above). They 
identified 
observational cycles of 56 and 64 years that seemed to underlie the data 
recorded on the tablet and, as a result, were able to propose a series 
of 
alternative chronologies. The three most frequently cited, commonly 
referred to as "High," "Middle," and "Low," place the fall of Babylon in 
1651, 1595, and 1531 B.C., respectively. Even though the reliability of 
the 
Venus data has been seriously questioned, Mesopotamian scholars have 
generally utilized one or another of these chronologies. 

Vahe Gurzadyan has now shown that the 56- and 64-year cycles, which had 
the 
practical effect of limiting the number of viable chronologies, cannot 
be 
extrapolated from the Venus Tablet data. Instead, there is only an 
8-year 
cycle, which, theoretically at least, permits many alternative 
chronologies 
for the 5 centuries leading up to Babylon's fall. However, because 
Gurzadyan has also been able to identify and date a pair of lunar 
eclipses 
from the early part of this half-millennium span, the fall of Babylon 
can 
be confidently pegged at 1499 B.C.

\end{document}